\documentclass{aa} 

\usepackage{graphicx}
\usepackage{txfonts}
\usepackage{hyperref}
\hypersetup{debug=true}
\hypersetup{
    colorlinks=true,
    linkcolor=blue,
    citecolor=blue,
    filecolor=magenta,      
    urlcolor=blue,
    pdftitle={Ambrosone+,2025},
    pdfpagemode=FullScreen,
    }

\usepackage[mathlines,switch]{lineno}

\begin{document}

\title{The origin of very high-energy diffuse $\gamma$-ray emission: The case for galactic source cocoons}
\titlerunning{Origin of  VHE diffuse $\gamma$-ray emission}
\authorrunning{Ambrosone et al.}

\author{Antonio Ambrosone\inst{1,2}\fnmsep\thanks{antonio.ambrosone@gssi.it} \and Carmelo Evoli\inst{1,2} \and Benedikt Schroer\inst{3} \and Pasquale Blasi\inst{1,2}}

\institute{Gran Sasso Science Institute (GSSI), Viale Francesco Crispi 7, 67100 L’Aquila, Italy \and INFN-Laboratori Nazionali del Gran Sasso(LNGS), via G. Acitelli 22, 67100 Assergi (AQ), Italy \and Department of Astronomy \& Astrophysics, University of Chicago, 5640 S Ellis Ave, Chicago, IL 60637, USA}

   \date{}

 
\abstract{\\\\The secondary/primary cosmic-ray ratios and the diffuse backgrounds of gamma rays and neutrinos provide us with complementary information about the transport of Galactic cosmic rays~(CRs). We used the recent measurement of the diffuse gamma ray background in the $\sim \rm TeV -\rm PeV$ range by LHAASO and of the very high-energy diffuse neutrino background from the Galactic  disc by IceCube to show that CRs may be  accumulating an approximately energy independent grammage $X\sim 0.4\, \rm g \, \rm cm^{-2}$, in regions where gamma rays and neutrinos are produced with a hard spectrum, resembling the source spectrum. We speculate that this grammage reflects the early stages of cosmic ray transport around sources, in what are referred to as cocoons, where particles spend $\sim 0.3\, \rm Myr$ before starting their journey in the Galactic environment.}


   \maketitle
%

\section{Introduction}

 The measurement of the spectra of primary and secondary, stable and unstable, cosmic-ray (CR) nuclei and of the diffuse fluxes of gamma rays and neutrinos represents an invaluable source of information about the origin of CRs, especially with regard to their transport in the Galaxy~\citep{Strong:2007nh,Gabici:2019jvz}.
Several experiments have been measuring the CR flux for different species with an accuracy that allows us to identify spectral features, such as a hardening at a rigidity of $\sim 300\, \rm GV$~\citep{PAMELA:2011mvy,AMS:2015tnn} and a softening at tens of TV~\citep{DAMPE:2019gys,CALET:2019bmh}.
In particular, the measurement of the B/C ratio and other similar secondary/primary ratios allowed us to conclude that the hardening at a few hundred GV is due to a change in the energy dependence of the diffusion coefficient~\citep{Genolini:2017qsj,AMS:2018tbl,CALET:2022dta,DAMPE:2022vwu,DiMauro:2023jgg,Ma:2022iji}, while the origin of the feature at $\sim 20$ TV is still subject of debate. The accuracy of direct measurements in the region below a few tens of TeV is not matched in the higher-energy region, where indirect experiments report somewhat different spectra of elements in the region of the knee, which is of the utmost importance for obvious reasons (see e.g. \citealt{KASCADEGrande:2017gtn} and \citealt{IceCube:2019hmk} for the proton spectrum measured by Kascade-GRANDE and IceCube-IceTop, respectively). Although local observations largely support the standard view that cosmic-ray transport proceeds by advection at low energies and diffusion at higher energies, the measured diffuse gamma-ray background, arising from inelastic collisions of cosmic rays with interstellar gas along the line of sight, still exhibits notable inconsistencies. The flux of diffuse gamma-ray emission (DGE) as measured by Fermi-LAT in the galactic disc~\citep{Fermi-LAT:2012edv}
shows that the spectrum from the region of the inner Galaxy is appreciably harder than expected, based on the CR fluxs at the solar neighbourhood~\citep{Gaggero:2014xla,Yang:2016jda,Fermi-LAT:2016zaq,Pothast:2018bvh}. This effect appears to persist at very high energies (\(\mathrm{TeV}\text{--}\mathrm{PeV}\)), as indicated by recent measurements from Tibet-AS\(\gamma\) \citep{TibetASgamma:2021tpz} and LHAASO \citep{LHAASO:2023gne,LHAASO:2024lnz}, prompting several studies to quantify the tension between these recent observations and the gamma-ray flux expected under a homogeneous advection-diffusion model~\citep{Gaggero:2017jts,Lipari:2018gzn,Cataldo:2020qla,Schwefer:2022zly,Vecchiotti:2024kkz}. There have been speculations concerning the existence of an unspecified additional component to the truly diffuse flux, solely aimed at fitting Fermi-LAT and LHAASO gamma-ray spectra~\citep{Zhang:2023ajh,Yang:2024igs}. Other authors have highlighted the possibility that the DGE may receive a contribution from a population of unresolved leptonic sources such as pulsar wind nebulae and/or TeV halos~\citep{Sudoh:2021avj,Martin:2022aun,Vecchiotti:2021vxp,Fang:2023ffx,Yan:2023hpt,Chen:2024yin}.  An important bit of information has been added with the recent measurement of a diffuse neutrino background by IceCube~\citep{IceCube:2023ame}, plausibly associated with the Galactic disc region, severely constraining the contribution of leptons to the DGE. The harder flux of gamma rays observed by Fermi-LAT in the direction of the inner Galaxy has  led some authors to speculate that different CR transport could be occurring in that region~\citep{Gaggero:2015xza,Gaggero:2017jts,Pagliaroli:2016lgg,Lipari:2018gzn,DeLaTorreLuque:2025zsv}. In non-linear theories of CR transport, this effect is caused by the fact that in the inner Galaxy CRs generate waves more effectively, and as a consequence transport becomes dominated by advection up to higher energies than in the outer Galaxy~\citep{Recchia:2016bnd}. However, in this case the hardening in the DGE should not extend to the $\gtrsim$ TeV energies measured by LHAASO. All these studies highlight that the observed DGE and high-energy neutrino flux challenge the conventional model.

It has been proposed that at some relatively high energies, $\gtrsim$ TeV, the contribution of the grammage traversed by CRs inside the accelerators (source grammage) should appear in the B/C and similar ratios: \cite{Aloisio:2015rsa} estimated a source grammage of order $\sim 0.2~\rm g/cm^2$ in a typical supernova remnant (SNR), while the effect of such source grammage and reacceleration at SNR shocks on secondary nuclei and antiprotons was calculated by~\cite{Blasi:2017caw,Bresci:2019aww}. All of these effects, which exhibit an energy dependence harder than the conventional diffusion expected in interstellar turbulence, are anticipated to become significant at high energies. Cocoons of reduced diffusivity are also predicted in non-linear theories of CR transport around sources~\citep{DAngelo:2015cfw,Nava:2016szf,Recchia:2021vfw,Schroer:2020dqy,Jacobs:2021qvh,Bao:2024esg}. 
More recently, \cite{Sun:2023ibg} and~\cite{Yang:2024igs} discussed the possibility that regions around CR sources could contribute a constant grammage that accommodates both the measured B/C ratio and the DGE observed by LHAASO in a unified picture in which the grammage accumulated on Galactic scales has a power-law behaviour. Combining a source grammage of $\sim 0.4~\rm g/cm^2$ with the standard model with a break in the diffusion coefficient around a few hundred GV rigidity provides a good fit to available data, as shown  by~\cite{Evoli:2019wwu}. However, this is not the case if the hardening in the B/C is only due to the source grammage, as assumed by~\cite{Sun:2023ibg} and~\cite{Yang:2024igs}. In this last scenario, one should not expect a corresponding break in the spectra of primary nuclei (e.g. H and He), unless such a break is imposed by hand at the source. Here we discuss the full implications of an energy-independent source grammage, tentatively indicated by the possible flattening of the high-energy B/C ratio in DAMPE and CALET data, for both the Galactic diffuse gamma-ray emission and the diffuse neutrino flux, and we comment on possible mechanisms by which such a grammage could be realized in nature.


\section{Theoretical framework}

There are different physical realizations of an energy independent grammage in the Galaxy, with correspondingly different signatures. A cocoon with reduced diffusivity around sources can be generated as a result of the non-linear feedback of accelerated particles~\citep{DAngelo:2015cfw,Nava:2016szf,Bao:2024esg,Schroer:2020dqy}, but typically in this case the grammage retains an energy dependence that is determined by the physical processes shaping the escape of CRs from the sources. An approximately constant grammage is accumulated in star clusters, at least up to $\sim 10$ TeV, as a result of the advection of CRs with the shocked collective wind of the cluster~\citep{Blasi:2023quw,Blasi2025}.

\begin{figure}[t!]
\includegraphics[width=\linewidth]{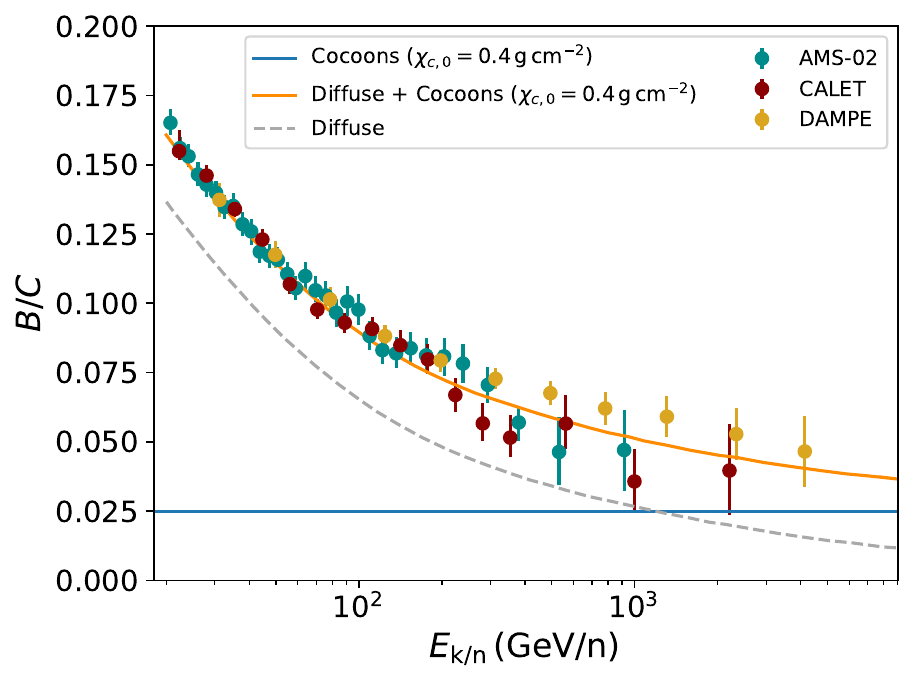}
\caption{B/C as a function of the kinetic energy per nucleon. The solid orange line represents the best fit  obtained by fitting AMS-02 measurements~\citep{AMS:2018tbl} (shown as cyan circles). The grey dashed line represents the Galactic grammage, while the solid blue line represents the cocoons contribution with $\chi_{c,0} = 0.4\, \rm g\, \rm cm^{-2}$. The data from CALET (in red)~\citep{CALET:2022dta} and DAMPE (in gold)~\citep{DAMPE:2022vwu}
are also shown.}
\label{fig:B_over_C}
\end{figure}

The grammage accumulated by CRs inside the parent SNR has also been estimated to be roughly energy independent up to $\sim 1$ TeV (the maximum energy at the end of the Sedov phase) and of order $0.2~\rm g/cm^2$ in~\cite{Aloisio:2015rsa}, with obvious uncertainties associated to the type of SNR and the environment in which the supernova occurs. In the absence of a definitive model for the cocoons, but having in mind star clusters as the most likely realization, below we assume that the cocoons are distributed in the disc of the Galaxy with a spatial distribution that reflects that of SNRs:
\begin{equation}
\rho(r,z) \propto \left(\frac{r}{r_{\odot}}\right)^{\alpha_s} e^{-\beta (r/r_{\odot}-1)} e^{-\frac{|z|}{z_{\rm 0}}}.
\end{equation}
Here $\rho(r,z)$ is the rate per unit volume, $r$ is the distance from the Galactic  centre, \( r_{\odot} = 8.5 \) kpc,  $z$ is the height with respect to the galactic plane, and $\alpha_s = 1.09$, $\beta = 3.87$, and $z_0 = 83\, \rm pc$ are fitting parameters, as derived by~\cite{Green:2015isa}. 
The normalisation is set by posing
\begin{equation}\label{eq:spatial_norm}
\int dV \rho(r,z) = \mathcal R_{\rm SN},
\end{equation}
where $\mathcal R_{\rm SN} = 1/30\, \rm yr^{-1}$ is the rate of supernova (SN) explosions. Finally, we consider a cylindrical disc for the Galaxy, with radius $R_{\rm G} = 10\,\rm kpc$ and half-height $h = 100\, \rm pc$. The grammage accumulated in cocoons is assumed to add to the standard Galactic grammage, which can be constrained by secondary-to-primary cosmic-ray ratios, as described by \cite{Evoli:2019wwu}. To compute this contribution we assume that each source of CRs is surrounded by a cocoon and contributes a CR spectrum $\mathcal{Q}_{ \rm CR}(p, r, z) = Q_{\rm CR}(p) \rho(r, z)$. 
$Q(E)$ is normalised such that
\begin{equation}\label{eq:norm_energy}
4\pi \int_{0}^{\infty} \! dp \, p^2 Q_{\rm CR} (p) T(p) = \epsilon E_{\rm SN} \, ,
\end{equation}
where
\begin{equation}
Q_{\rm CR} (p) \propto \left(\frac{p}{m_{\alpha} c}\right)^{-\gamma} e^{-\frac{p}{p_{\rm max}}} \, ,
\end{equation}
and $T(p)$ is the kinetic energy of a CR particle of mass \( m_\alpha \),  $c$ is the speed of light, and $p_{\rm max}$ is a high-energy cut-off. We adopt $E_{\rm SN} = 10^{51}\, \rm erg$ as the kinetic energy released by SNRs. Consistent with the best fit to direct cosmic-ray measurements reported by~\cite{Evoli:2019wwu}, we set the proton injection efficiency to \(\epsilon = 5\%\) and the spectral index to \(\gamma = 4.26\). For helium, a lower efficiency of \(\epsilon = 1.2\%\), but a slightly harder spectral index of \(\gamma = 4.16\) is required. The maximum energy is assumed to extend up to the cosmic-ray knee, with \( p_{\rm max} \simeq Z\,\text{PeV} \).

\begin{figure*}[t!]
\centering
\includegraphics[width=0.49\linewidth]{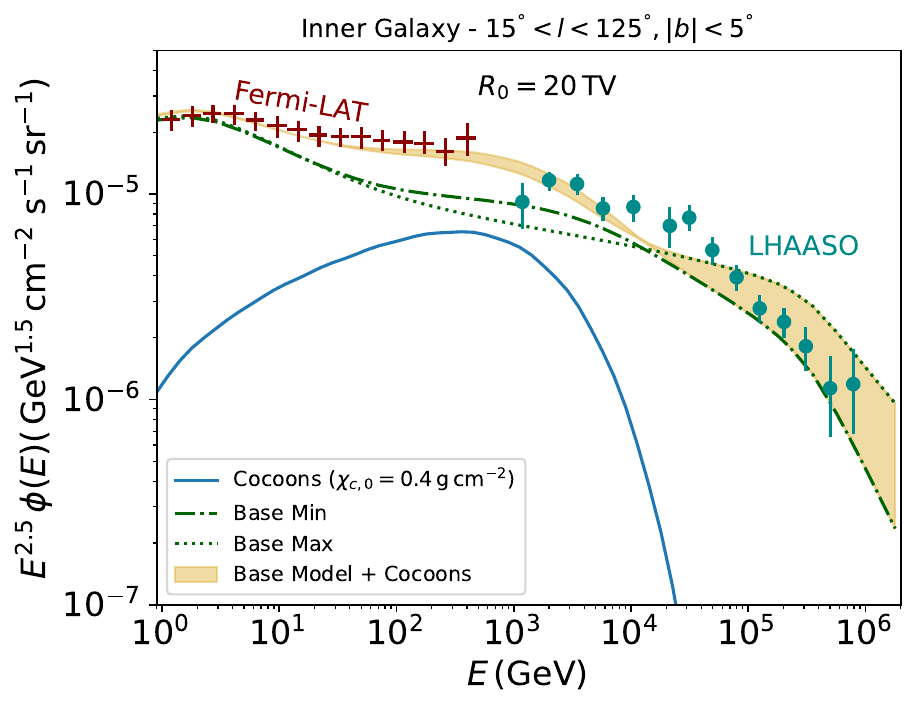}
\includegraphics[width=0.49\linewidth]{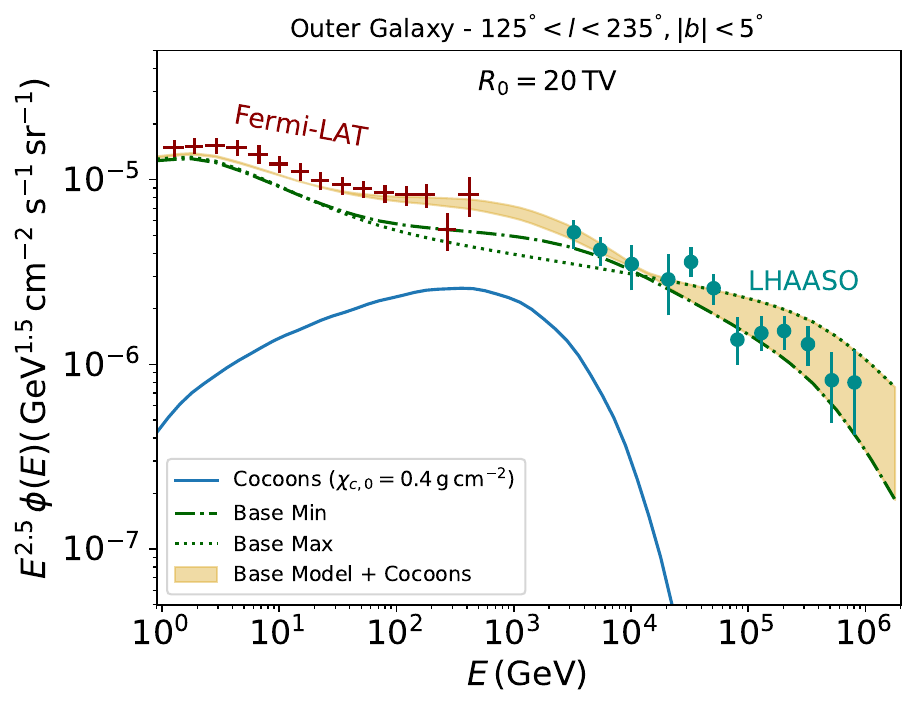}
\caption{\textbf{Left}: Inner galaxy gamma-ray SEDs as a function of the energy for the DGE + cocoons contributions (golden band) and for the cocoons only (blue line).  We rescale the predicted SEDs in order to account for the LHAASO mask in the galactic disc. \textbf{Right}: Same as left panel,  but for the outer galaxy. In both panels, the gamma-ray SEDs are compared with  Fermi-LAT and LHAASO  data reported by~\cite{LHAASO:2024lnz}.  }
\label{fig:inner_galaxy_gamma}
\end{figure*}
The local density of CRs in a generic point where a cocoon is present and particles are confined for a time $\tau_{\rm c} (p)$, can be estimated as
\begin{equation}\label{eq:transport_disk}
f_{\rm CR} (p,r,z) = \mathcal Q_{ \rm CR}(p,r,z) \tau_{\rm c} (p) =  Q_{ \rm CR}(p) \rho(r,z) \frac{\chi_c(p)}{m_p n_{\rm gas} c},
\end{equation}
where $n_{\rm gas}$ is the gas density in the cocoon, and we introduced the grammage of the cocoon as $\chi_c(p)$. 
Equation~\ref{eq:transport_disk} holds if escape from the cocoons occurs faster than energy losses. If the cocoons are identified as star clusters, this is a reasonable approximation as long as the gas density in the cluster is appreciably lower than a few particles per $\rm cm^3$, although higher densities are sometimes invoked \citep{Blasi:2023quw,Blasi2025}. 
The assumption of neglecting energy losses is definitely justified if the cocoons are interpreted as the result of confinement inside SNRs.
As discussed above, generally, $\chi_c$ is roughly energy independent up to some rigidity  \( R_0 \).
In the case of stellar clusters, this dependence would stem from the fact that at low rigidities (typically below $\sim 10$ TeV) escape is advection dominated \citep{Blasi:2023quw,Menchiari:2024uce}.
On the other hand, at SNR shocks, particles with rigidity below the maximum rigidity at the end of the Sedov-Taylor phase ($R_0$) are trapped inside the remnant, while the higher-energy particles escape at earlier times. Hence, in this scenario particles with rigidity $\lesssim R_0$ experience the same grammage \citep{Aloisio:2015rsa}.
For simplicity and to avoid dependence on details of a specific realization of the cocoons, we assume the following scaling:
\begin{equation}\label{eq:confinement_time}
\chi_c (R) = \chi_{c,0} \, e^{-\frac{R}{R_0}} \,.
\end{equation}
Here \( R_0 \sim 20 \)~TV, as this choice provides a good match to the gamma-ray data; further discussion on the impact of different \( R_0 \) values is provided in the appendix. At least for the case of star clusters, one expects that the transition from advective to diffusive escape is reflected in a change in grammage from constant in energy to a power law~\citep{Blasi:2023quw,Menchiari:2024uce}. These details may well change (and in fact improve) the agreement with data in the same energy range (see appendix for a more quantitative discussion). The gamma rays and neutrino production rates at a given point in the galactic disc, where cocoons are assumed to be located, reads 
\begin{equation}\label{eq:prod_rate_in_given_point}
\begin{split}
\mathcal{Q}_{\gamma,\nu}(E,r,z) & = \frac{\rho(r,z)}{m_p} \sum_{i = \text{H},\text{He}} \int_{E_{\rm th,i}}^{+\infty} dE_i Q_{\rm CR, i}^{E} \left(E_i\right) \chi_c \left(E_i\right) \times  \\
& \times \frac{d\sigma^{i}_{\gamma, \nu}}{dE_i}(E_i,E) = \rho(r,z) Q_{\gamma, \nu}(E), 
\end{split}
\end{equation}
where $Q_{\rm CR}^{E}(E) = 4\pi p^2  Q_{\rm CR}(p(E)) \frac{dp}{dE}$, and $\frac{d\sigma^{i}_{\gamma, \nu}}{dE_i}(E_i,E)$ is the differential cross-section for producing either gamma rays or neutrinos for a given primary nucleus. We computed the differential cross-sections by interpolating the model provided by~\citet{Koldobskiy:2021nld} and publicly available in  \url{github.com/aafragpy/aafragpy}. We   verified that using other cross-sections such as that reported by~\citet{Orusa:2023bnl,Kelner:2006tc,Joshi:2013aua} does not   appreciably change our results. The systematic uncertainties attributable to this ingredient is $\lesssim 15\%$ above $E_{\gamma} \sim 1\, \rm GeV$. Finally, the angle-integrated flux from cocoon contributions reads
\begin{equation}\label{eq:cocoon_contribution}
\Phi^{\rm cocoon}_{\gamma/\nu}(E) = \frac{Q_{\gamma/\nu}(E)}{4\pi}\int d \Omega \int_0^\infty ds \, \rho(s,\Omega),
\end{equation}
where  $s$ is the distance to the solar position along the line of sight and $\Omega$ is the solid angle. Given that \(\chi_c\) is too small to alter the spectrum released in the ISM, primary CRs produced by the sources leave the cocoons essentially unchanged and generate the standard DGE by interacting with interstellar gas. To compare our model to a  reference DGE model, we show the result of~\cite{Luque:2022buq}, which solves the homogeneous CR transport equation using the DRAGON2 code~\citep{Evoli:2016xgn,Evoli:2017vim} and normalizes the results to the local CR flux. In order to accommodate the discordant proton spectrum measurements by KASCADE-Grande and IceTop at high energy, that work considers two source-spectrum models, labelled \textit{base min} and \textit{base max}. The source and injection parameters are fully detailed in~\citet{Luque:2022buq}. The gamma-ray and neutrino diffuse fluxes are computed using the HERMES code~\citep{Dundovic:2021ryb}, which integrates, along each line of sight, the gamma-ray and neutrino emissivity produced by hadronic collisions with the ISM. In the following we show the total secondary flux expected at Earth as $\Phi_{\gamma/\nu}^{\rm tot}(E) = \Phi_{\gamma/\nu}^{\rm cocoons}(E) + \Phi_{\gamma/\nu}^{\rm DGE}(E) $, with $\Phi_{\gamma/\nu}^{\rm DGE}(E)$ being the diffuse emission.

\section{Results}

We calculated the boron-to-carbon (B/C) ratio resulting from a combination of CR transport in the Galaxy and in the cocoons, following~\cite{Schroer:2021ojh,Evoli:2019wwu}, inserting an injection term into the secondary CRs transport equation. We find that a good fit is achieved with a non-zero $\chi_{c,0}$ in the range of $[0.3-0.6]\, \rm g\, \rm cm^{-2}$, with a best-fit value $\chi_{c,0} = 0.4\, \rm cm^{-2}$, which we take as a fiducial value for the following results. The resulting best-fit ratio as a function of the kinetic energy per nucleon is shown in Fig.~\ref{fig:B_over_C}.
 Part of the hardening that can be found above $300\,$~GeV/n is due to the change in the Galactic diffusion coefficient necessary to interpret the spectral break in the primary nuclei. Part of it is due to the contribution of the grammage accumulated by CRs in the cocoons, which is responsible for the high-energy flattening and for slightly reducing the contribution of the galactic grammage.
The inner and outer galaxy gamma-ray spectral energy distributions (SEDs) are shown in Fig.~\ref{fig:inner_galaxy_gamma}~(see appendix for the results for the neutrinos). For the inner galaxy, the SED normalisations are scaled by a factor of 65\% to account for the LHAASO mask over the galactic disc. \cite{Zhang:2023ajh} shows that the mask changes only the normalization of the Fermi-LAT data, and for this reason we rescaled the theoretical predictions by the same factor. The contribution of the cocoons allows us to obtain a satisfactory description of the observed gamma-ray diffuse emission  from the inner and the outer Galaxy. We note that the predicted flux of gamma rays from the cocoons in the range around $\sim 10$ TeV depends on the details of particle escape from the cocoons, which for simplicity we modelled as an exponential loss of confinement. Finally, we note that the contribution of the cocoons to the diffuse fluxes of gamma rays relies upon the assumptions that the lifetime of a cocoon exceeds the confinement time necessary for CRs to accumulate the grammage $
\chi_c$. This condition reads
\begin{equation}\label{eq:tau_life}
\tau_{\rm life} \gtrsim 2.5 \times 10^{5} \, \rm yr \,  \left(\frac{n_{\rm gas}}{1\, \rm cm^{-3}}\right)^{-1} \left( \frac{\chi_{c,0}}{0.4 \, \rm g \,  \rm cm^{-2}} \right)~.
\end{equation}
In addition, for the cocoons to contribute to the diffuse background and not be detectable as individual gamma ray sources, it is necessary that they fill a relatively large fraction of the volume of the disc without exceeding it, a condition that leads to a constraint on the size of a cocoon,
\begin{equation}
R_c \lesssim 120 \, \rm pc \left(\frac{\chi_{c,0}}{0.4\, \rm g \, \rm cm^{-2}}\right)^{-1/3}  \left(\frac{n_{\rm gas}}{1\, \rm cm^{-3}}\right)^{1/3},
\end{equation}
a condition that is easily fulfilled, for instance by star clusters.
\section{Conclusions}
Local measurements of CR primaries and secondaries now clearly indicate that CR transport at energies \(E \gtrsim 1\) TeV differs significantly from lower energies~\citep{Evoli:2023kxd}. Whether this transition arises from different scattering mechanisms, for instance self-generated versus pre-existing turbulence~\citep{Blasi:2012yr}, or from a non-trivial spatial variation in the diffusion coefficient~\citep{Tomassetti:2012ga,Recchia:2023mmg} remains unresolved. Crucially, measurements of diffuse gamma-ray emission from the Galactic disc offer a key opportunity to probe CR distributions in situ across extended regions of the Galaxy, thus helping to break degeneracies present in purely local CR data. In recent years, the diffuse gamma-ray flux in the TeV to PeV range has been measured in various regions of the Galactic plane by Milagro~\citep{Abdo:2008if}, H.E.S.S.~\citep{HESS:2014ree}, ARGO-YBJ~\citep{ARGO-YBJ:2015cpa}, HAWC~\citep{HAWC:2023wdq}, Tibet AS\(\gamma\)~\citep{TibetASgamma:2021tpz}, and LHAASO~\citep{LHAASO:2024lnz}. Alongside the hadronically generated emission observed by the Fermi-LAT in the $\sim$GeV domain~\citep{Fermi-LAT:2012edv}, these new data extend our view of Galactic gamma-ray production up to \(\lesssim\)~PeV. A major challenge at these energies, beyond separating gamma rays from charged-particle contamination, is to distinguish truly interstellar emission from the integrated flux of unresolved sources. Depending on model assumptions, estimates diverge on how much of the diffuse flux may stem from faint pulsar wind nebulae or TeV halos that lie below TeV detection thresholds~\citep{Steppa:2020qwe,TibetASgamma:2021tpz,Vecchiotti:2021vxp,Lipari:2024pzo}. The recent detection of the neutrino counterpart of the DGE is critical to confirming its hadronic origin~\citep{IceCube:2023ame}. Despite some discrepancies across instruments and ongoing uncertainties in constraining the fraction of unresolved sources, there is however robust evidence that TeV-PeV gamma-ray measurements often exceed predictions based on the local cosmic-ray flux alone. Observed departures appear to require variations in cosmic-ray densities and spectra across the Galactic disc beyond what standard transport models would predict. These findings might underscore that our understanding of Galactic CR transport and its microphysical foundations remains incomplete. In this letter we introduced a slight modification to the standard scenario in which the overall behaviour of primary nuclei is governed by an effective diffusion coefficient whose energy dependence drives the dominant transport process, while an additional grammage of \(\sim 0.4\,\mathrm{g\,cm^{-2}}\) is accumulated in or around the CR sources themselves. This extended cocoon scenario naturally addresses both the tentative flattening of secondary-to-primary ratios at high energies and the mismatch in the diffuse gamma-ray background. 
Unlike previous studies (e.g.~\citealt{Gaggero:2015xza,Gaggero:2017jts}), which attribute the hardening of the gamma-ray spectrum towards the inner Galaxy to a spatially varying cosmic-ray diffusion slope, our model naturally reproduces this feature through the increased density of CR cocoons,   each with a harder intrinsic spectrum,   near the Galactic centre, reflecting the underlying SNR distribution. We highlight that the most plausible manifestation of such cocoons may be star-cluster wind bubbles, where CRs below \(\sim 10\) TeV escape mainly via advection, while higher-energy particles diffuse out~\citep{Blasi:2023quw}. Consequently, the accumulated grammage remains roughly energy independent up to \(\sim 10\) TeV and then transitions to a steeper, diffusion-driven dependence. For simplicity, our calculations adopt an exponential suppression at high energies; thus, the high-energy cocoon contribution we obtain should be taken as a conservative (lower) limit. In the Appendix we explore how alternative energy dependences  of the cocoon grammage could affect these conclusions. It is worth noting that other cocoon models, such as in~\cite{Yang:2024igs}, or the Galactic transport scenario of~\cite{Recchia:2023mmg} likewise predict an energy-independent grammage. However, in the former the break in the primary spectrum (coincident with the B/C ratio hardening) must be artificially imposed on the source spectrum, a somewhat contrived coincidence. Meanwhile, the latter yields a generally hard high-energy gamma-ray spectrum, but does not naturally distinguish between the inner and outer Galaxy, in contrast to what we find here.

\begin{acknowledgements}
     We thank the authors of~\cite{Luque:2022buq} for providing the gamma-ray and neutrino diffuse maps of their \texttt{base} model, available in the repository~\url{https://zenodo.org/records/12802088}.
AA acknowledges the support of the project ``NUSES - A pathfinder for studying astrophysical neutrinos and electromagnetic signals of seismic origin from space'' (Cod.~id.~Ugov: NUSES; CUP: D12F19000040003). The work of PB and CE has been partially funded by the European Union – Next Generation EU, through PRIN-MUR 2022TJW4EJ and by the European Union – NextGenerationEU under the MUR National Innovation Ecosystem grant ECS00000041 – VITALITY/ASTRA – CUP D13C21000430001. 
The work of BS was supported by NSF through grants NSF-2009326, and NSF-2010240.
\end{acknowledgements}

\bibliographystyle{aa}
\bibliography{biblio}

\begin{thebibliography}{75}
\expandafter\ifx\csname natexlab\endcsname\relax\def\natexlab#1{#1}\fi

\bibitem[{Aartsen {et~al.}(2019)}]{IceCube:2019hmk}
Aartsen, M.~G. {et~al.} 2019, Phys. Rev. D, 100, 082002

\bibitem[{Abbasi {et~al.}(2023)}]{IceCube:2023ame}
Abbasi, R. {et~al.} 2023, Science, 380, adc9818

\bibitem[{Abdo {et~al.}(2008)}]{Abdo:2008if}
Abdo, A.~A. {et~al.} 2008, Astrophys. J., 688, 1078

\bibitem[{Abramowski {et~al.}(2014)}]{HESS:2014ree}
Abramowski, A. {et~al.} 2014, Phys. Rev. D, 90, 122007

\bibitem[{Acero {et~al.}(2016)}]{Fermi-LAT:2016zaq}
Acero, F. {et~al.} 2016, Astrophys. J. Suppl., 223, 26

\bibitem[{Ackermann {et~al.}(2012)}]{Fermi-LAT:2012edv}
Ackermann, M. {et~al.} 2012, Astrophys. J., 750, 3

\bibitem[{Adriani {et~al.}(2011)}]{PAMELA:2011mvy}
Adriani, O. {et~al.} 2011, Science, 332, 69

\bibitem[{Adriani {et~al.}(2019)}]{CALET:2019bmh}
Adriani, O. {et~al.} 2019, Phys. Rev. Lett., 122, 181102

\bibitem[{Adriani {et~al.}(2022)}]{CALET:2022dta}
Adriani, O. {et~al.} 2022, Phys. Rev. Lett., 129, 251103

\bibitem[{Aguilar {et~al.}(2015)}]{AMS:2015tnn}
Aguilar, M. {et~al.} 2015, Phys. Rev. Lett., 114, 171103

\bibitem[{Aguilar {et~al.}(2018)}]{AMS:2018tbl}
Aguilar, M. {et~al.} 2018, Phys. Rev. Lett., 120, 021101

\bibitem[{Alfaro {et~al.}(2024)}]{HAWC:2023wdq}
Alfaro, R. {et~al.} 2024, Astrophys. J., 961, 104

\bibitem[{Aloisio {et~al.}(2015)Aloisio, Blasi, \& Serpico}]{Aloisio:2015rsa}
Aloisio, R., Blasi, P., \& Serpico, P. 2015, Astron. Astrophys., 583, A95

\bibitem[{Ambrosone {et~al.}(2024)Ambrosone, Groth, Peretti, \&
  Ahlers}]{Ambrosone:2023hsz}
Ambrosone, A., Groth, K.~M., Peretti, E., \& Ahlers, M. 2024, Phys. Rev. D,
  109, 043007

\bibitem[{Amenomori {et~al.}(2021)}]{TibetASgamma:2021tpz}
Amenomori, M. {et~al.} 2021, Phys. Rev. Lett., 126, 141101

\bibitem[{An {et~al.}(2019)}]{DAMPE:2019gys}
An, Q. {et~al.} 2019, Sci. Adv., 5, eaax3793

\bibitem[{Arteaga-Vel\'azquez {et~al.}(2018)}]{KASCADEGrande:2017gtn}
Arteaga-Vel\'azquez, C.~J. {et~al.} 2018, PoS, ICRC2017, 316

\bibitem[{Bao {et~al.}(2024)Bao, Blasi, \& Chen}]{Bao:2024esg}
Bao, Y., Blasi, P., \& Chen, Y. 2024, Astrophys. J., 966, 224

\bibitem[{Bartoli {et~al.}(2015)}]{ARGO-YBJ:2015cpa}
Bartoli, B. {et~al.} 2015, Astrophys. J., 806, 20

\bibitem[{Blasi(2017)}]{Blasi:2017caw}
Blasi, P. 2017, Mon. Not. Roy. Astron. Soc., 471, 1662

\bibitem[{{Blasi}(2025)}]{Blasi2025}
{Blasi}, P. 2025, \aap, 694, A244

\bibitem[{Blasi {et~al.}(2012)Blasi, Amato, \& Serpico}]{Blasi:2012yr}
Blasi, P., Amato, E., \& Serpico, P.~D. 2012, Phys. Rev. Lett., 109, 061101

\bibitem[{Blasi \& Morlino(2023)}]{Blasi:2023quw}
Blasi, P. \& Morlino, G. 2023, Mon. Not. Roy. Astron. Soc., 523, 4015

\bibitem[{Bresci {et~al.}(2019)Bresci, Amato, Blasi, \&
  Morlino}]{Bresci:2019aww}
Bresci, V., Amato, E., Blasi, P., \& Morlino, G. 2019, Mon. Not. Roy. Astron.
  Soc., 488, 2068

\bibitem[{Cao {et~al.}(2023)}]{LHAASO:2023gne}
Cao, Z. {et~al.} 2023, Phys. Rev. Lett., 131, 151001

\bibitem[{Cao {et~al.}(2025)}]{LHAASO:2024lnz}
Cao, Z. {et~al.} 2025, Phys. Rev. Lett., 134, 081002

\bibitem[{Cataldo {et~al.}(2020)Cataldo, Pagliaroli, Vecchiotti, \&
  Villante}]{Cataldo:2020qla}
Cataldo, M., Pagliaroli, G., Vecchiotti, V., \& Villante, F.~L. 2020,
  Astrophys. J., 904, 85

\bibitem[{Chen {et~al.}(2024)Chen, Fang, \& Bi}]{Chen:2024yin}
Chen, E.-S., Fang, K., \& Bi, X.-J. 2024, Chin. Phys. C, 48, 115105

\bibitem[{{DAMPE Collaboration}(2022)}]{DAMPE:2022vwu}
{DAMPE Collaboration}. 2022, Sci. Bull., 67, 2162

\bibitem[{D'Angelo {et~al.}(2016)D'Angelo, Blasi, \& Amato}]{DAngelo:2015cfw}
D'Angelo, M., Blasi, P., \& Amato, E. 2016, Phys. Rev. D, 94, 083003

\bibitem[{{De la Torre Luque} {et~al.}(2023){De la Torre Luque}, Gaggero,
  Grasso, Fornieri, Egberts, Steppa, \& Evoli}]{Luque:2022buq}
{De la Torre Luque}, P., Gaggero, D., Grasso, D., {et~al.} 2023, Astron.
  Astrophys., 672, A58

\bibitem[{{De la Torre Luque} {et~al.}(2025){De la Torre Luque}, Gaggero,
  Grasso, Marinelli, \& Rocamora}]{DeLaTorreLuque:2025zsv}
{De la Torre Luque}, P., Gaggero, D., Grasso, D., Marinelli, A., \& Rocamora,
  M. 2025 [\eprint[arXiv]{2502.18268}]

\bibitem[{Di~Mauro {et~al.}(2024)Di~Mauro, Korsmeier, \&
  Cuoco}]{DiMauro:2023jgg}
Di~Mauro, M., Korsmeier, M., \& Cuoco, A. 2024, Phys. Rev. D, 109, 123003

\bibitem[{Dundovic {et~al.}(2021)Dundovic, Evoli, Gaggero, \&
  Grasso}]{Dundovic:2021ryb}
Dundovic, A., Evoli, C., Gaggero, D., \& Grasso, D. 2021, Astron. Astrophys.,
  653, A18

\bibitem[{Evoli {et~al.}(2019)Evoli, Aloisio, \& Blasi}]{Evoli:2019wwu}
Evoli, C., Aloisio, R., \& Blasi, P. 2019, Phys. Rev. D, 99, 103023

\bibitem[{Evoli \& Dupletsa(2024)}]{Evoli:2023kxd}
Evoli, C. \& Dupletsa, U. 2024, Proc. Int. Sch. Phys. Fermi, 208, 257

\bibitem[{Evoli {et~al.}(2017)Evoli, Gaggero, Vittino, Di~Bernardo, Di~Mauro,
  Ligorini, Ullio, \& Grasso}]{Evoli:2016xgn}
Evoli, C., Gaggero, D., Vittino, A., {et~al.} 2017, JCAP, 02, 015

\bibitem[{Evoli {et~al.}(2018)Evoli, Gaggero, Vittino, Di~Mauro, Grasso, \&
  Mazziotta}]{Evoli:2017vim}
Evoli, C., Gaggero, D., Vittino, A., {et~al.} 2018, JCAP, 07, 006

\bibitem[{Fang \& Murase(2023)}]{Fang:2023ffx}
Fang, K. \& Murase, K. 2023, Astrophys. J. Lett., 957, L6

\bibitem[{Gabici {et~al.}(2019)Gabici, Evoli, Gaggero, Lipari, Mertsch,
  Orlando, Strong, \& Vittino}]{Gabici:2019jvz}
Gabici, S., Evoli, C., Gaggero, D., {et~al.} 2019, Int. J. Mod. Phys. D, 28,
  1930022

\bibitem[{Gaggero {et~al.}(2017)Gaggero, Grasso, Marinelli, Taoso, \&
  Urbano}]{Gaggero:2017jts}
Gaggero, D., Grasso, D., Marinelli, A., Taoso, M., \& Urbano, A. 2017, Phys.
  Rev. Lett., 119, 031101

\bibitem[{Gaggero {et~al.}(2015{\natexlab{a}})Gaggero, Grasso, Marinelli,
  Urbano, \& Valli}]{Gaggero:2015xza}
Gaggero, D., Grasso, D., Marinelli, A., Urbano, A., \& Valli, M.
  2015{\natexlab{a}}, Astrophys. J. Lett., 815, L25

\bibitem[{Gaggero {et~al.}(2015{\natexlab{b}})Gaggero, Urbano, Valli, \&
  Ullio}]{Gaggero:2014xla}
Gaggero, D., Urbano, A., Valli, M., \& Ullio, P. 2015{\natexlab{b}}, Phys. Rev.
  D, 91, 083012

\bibitem[{Genolini {et~al.}(2018)}]{Genolini:2017qsj}
Genolini, Y. {et~al.} 2018, PoS, ICRC2017, 268

\bibitem[{Green(2015)}]{Green:2015isa}
Green, D.~A. 2015, Mon. Not. Roy. Astron. Soc., 454, 1517

\bibitem[{Jacobs {et~al.}(2022)Jacobs, Mertsch, \& Phan}]{Jacobs:2021qvh}
Jacobs, H., Mertsch, P., \& Phan, V. H.~M. 2022, JCAP, 05, 024

\bibitem[{Joshi {et~al.}(2014)Joshi, Winter, \& Gupta}]{Joshi:2013aua}
Joshi, J.~C., Winter, W., \& Gupta, N. 2014, Mon. Not. Roy. Astron. Soc., 439,
  3414, [Erratum: Mon.Not.Roy.Astron.Soc. 446, 892 (2014)]

\bibitem[{Kelner {et~al.}(2006)Kelner, Aharonian, \& Bugayov}]{Kelner:2006tc}
Kelner, S.~R., Aharonian, F.~A., \& Bugayov, V.~V. 2006, Phys. Rev. D, 74,
  034018, [Erratum: Phys.Rev.D 79, 039901 (2009)]

\bibitem[{Koldobskiy {et~al.}(2021)Koldobskiy, Kachelrie\ss{}, Lskavyan,
  Neronov, Ostapchenko, \& Semikoz}]{Koldobskiy:2021nld}
Koldobskiy, S., Kachelrie\ss{}, M., Lskavyan, A., {et~al.} 2021, Phys. Rev. D,
  104, 123027

\bibitem[{Lipari \& Vernetto(2018)}]{Lipari:2018gzn}
Lipari, P. \& Vernetto, S. 2018, Phys. Rev. D, 98, 043003

\bibitem[{Lipari \& Vernetto(2025)}]{Lipari:2024pzo}
Lipari, P. \& Vernetto, S. 2025, Phys. Rev. D, 111, 063035

\bibitem[{Ma {et~al.}(2023)Ma, Xu, Yuan, Bi, Fan, Moskalenko, \&
  Yue}]{Ma:2022iji}
Ma, P.-X., Xu, Z.-H., Yuan, Q., {et~al.} 2023, Front. Phys. (Beijing), 18,
  44301

\bibitem[{Martin {et~al.}(2022)Martin, Tibaldo, Marcowith, \&
  Abdollahi}]{Martin:2022aun}
Martin, P., Tibaldo, L., Marcowith, A., \& Abdollahi, S. 2022, Astron.
  Astrophys., 666, A7

\bibitem[{Menchiari {et~al.}(2025)Menchiari, Morlino, Amato, Bucciantini,
  Peron, \& Sacco}]{Menchiari:2024uce}
Menchiari, S., Morlino, G., Amato, E., {et~al.} 2025, Astron. Astrophys., 695,
  A175

\bibitem[{Nava {et~al.}(2016)Nava, Gabici, Marcowith, Morlino, \&
  Ptuskin}]{Nava:2016szf}
Nava, L., Gabici, S., Marcowith, A., Morlino, G., \& Ptuskin, V.~S. 2016, Mon.
  Not. Roy. Astron. Soc., 461, 3552

\bibitem[{Orusa {et~al.}(2023)Orusa, Di~Mauro, Donato, \&
  Korsmeier}]{Orusa:2023bnl}
Orusa, L., Di~Mauro, M., Donato, F., \& Korsmeier, M. 2023, Phys. Rev. D, 107,
  083031

\bibitem[{Pagliaroli {et~al.}(2016)Pagliaroli, Evoli, \&
  Villante}]{Pagliaroli:2016lgg}
Pagliaroli, G., Evoli, C., \& Villante, F.~L. 2016, JCAP, 11, 004

\bibitem[{Pothast {et~al.}(2018)Pothast, Gaggero, Storm, \&
  Weniger}]{Pothast:2018bvh}
Pothast, M., Gaggero, D., Storm, E., \& Weniger, C. 2018, JCAP, 10, 045

\bibitem[{Recchia {et~al.}(2016)Recchia, Blasi, \& Morlino}]{Recchia:2016bnd}
Recchia, S., Blasi, P., \& Morlino, G. 2016, Mon. Not. Roy. Astron. Soc., 462,
  L88

\bibitem[{Recchia \& Gabici(2024)}]{Recchia:2023mmg}
Recchia, S. \& Gabici, S. 2024, Astron. Astrophys., 692, A20

\bibitem[{Recchia {et~al.}(2022)Recchia, Galli, Nava, Padovani, Gabici,
  Marcowith, Ptuskin, \& Morlino}]{Recchia:2021vfw}
Recchia, S., Galli, D., Nava, L., {et~al.} 2022, Astron. Astrophys., 660, A57

\bibitem[{Schroer {et~al.}(2021{\natexlab{a}})Schroer, Evoli, \&
  Blasi}]{Schroer:2021ojh}
Schroer, B., Evoli, C., \& Blasi, P. 2021{\natexlab{a}}, Phys. Rev. D, 103,
  123010

\bibitem[{Schroer {et~al.}(2021{\natexlab{b}})Schroer, Pezzi, Caprioli,
  Haggerty, \& Blasi}]{Schroer:2020dqy}
Schroer, B., Pezzi, O., Caprioli, D., Haggerty, C., \& Blasi, P.
  2021{\natexlab{b}}, Astrophys. J. Lett., 914, L13

\bibitem[{Schwefer {et~al.}(2023)Schwefer, Mertsch, \&
  Wiebusch}]{Schwefer:2022zly}
Schwefer, G., Mertsch, P., \& Wiebusch, C. 2023, Astrophys. J., 949, 16

\bibitem[{Steppa \& Egberts(2020)}]{Steppa:2020qwe}
Steppa, C. \& Egberts, K. 2020, Astron. Astrophys., 643, A137

\bibitem[{Strong {et~al.}(2007)Strong, Moskalenko, \& Ptuskin}]{Strong:2007nh}
Strong, A.~W., Moskalenko, I.~V., \& Ptuskin, V.~S. 2007, Ann. Rev. Nucl. Part.
  Sci., 57, 285

\bibitem[{Sudoh {et~al.}(2021)Sudoh, Linden, \& Hooper}]{Sudoh:2021avj}
Sudoh, T., Linden, T., \& Hooper, D. 2021, JCAP, 08, 010

\bibitem[{Sun {et~al.}(2024)Sun, Zhang, Yuan, Liu, \& Guo}]{Sun:2023ibg}
Sun, D.-X., Zhang, P.-P., Yuan, Q., Liu, W., \& Guo, Y.-Q. 2024, Phys. Rev. D,
  110, 103039

\bibitem[{Tomassetti(2012)}]{Tomassetti:2012ga}
Tomassetti, N. 2012, Astrophys. J. Lett., 752, L13

\bibitem[{Vecchiotti {et~al.}(2022)Vecchiotti, Pagliaroli, \&
  Villante}]{Vecchiotti:2021vxp}
Vecchiotti, V., Pagliaroli, G., \& Villante, F.~L. 2022, Commun. Phys., 5, 161

\bibitem[{Vecchiotti {et~al.}(2024)Vecchiotti, Peron, Amato, Menchiari,
  Morlino, Pagliaroli, \& Villante}]{Vecchiotti:2024kkz}
Vecchiotti, V., Peron, G., Amato, E., {et~al.} 2024
  [\eprint[arXiv]{2411.11439}]

\bibitem[{Yan {et~al.}(2024)Yan, Liu, Zhang, Li, Yuan, \& Wang}]{Yan:2023hpt}
Yan, K., Liu, R.-Y., Zhang, R., {et~al.} 2024, Nature Astron., 8, 628

\bibitem[{{Yang} \& {Aharonian}(2025)}]{Yang:2024igs}
{Yang}, R. \& {Aharonian}, F. 2025, \prd, 111, 083040

\bibitem[{Yang {et~al.}(2016)Yang, Aharonian, \& Evoli}]{Yang:2016jda}
Yang, R., Aharonian, F., \& Evoli, C. 2016, Phys. Rev. D, 93, 123007

\bibitem[{Zhang {et~al.}(2023)Zhang, Huang, Xu, Zhao, \& Yuan}]{Zhang:2023ajh}
Zhang, R., Huang, X., Xu, Z.-H., Zhao, S., \& Yuan, Q. 2023, Astrophys. J.,
  957, 43

\end{thebibliography}

\begin{appendix}
    \section{Impact of the value of \texorpdfstring{$R_0$}{R0} on the cocoon contribution to the diffuse background}\label{app:cutoff}

\begin{figure*}[h!]
\centering
\includegraphics[width=0.47\linewidth]{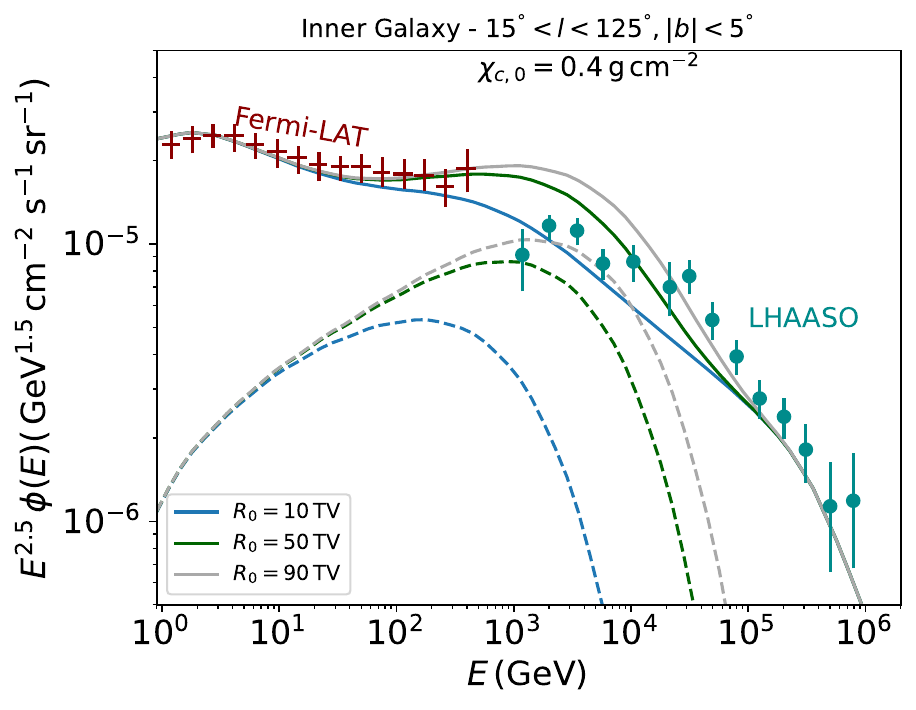}
\includegraphics[width=0.47\linewidth]{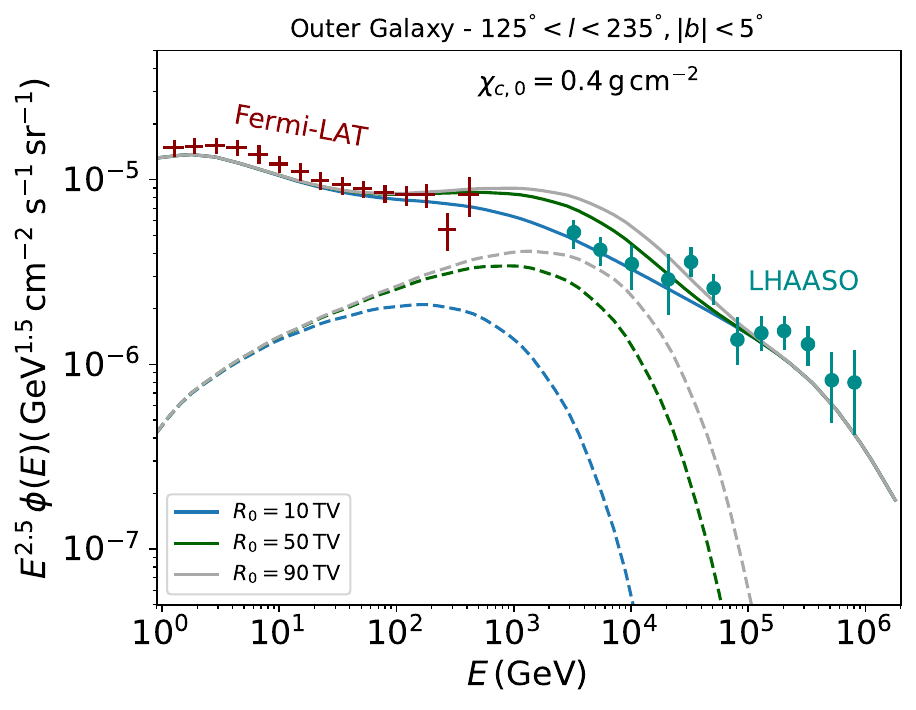}
\caption{\textbf{Left:} Gamma-ray SED of the cocoons in the inner Galaxy for several cut-off values, with \(\chi_{c,0} = 0.4\,\mathrm{g\,cm^{-2}}\) (dashed lines). The blue, green, and grey curves correspond to \(R_0 = 10\,\mathrm{TV}\), \(R_0 = 50\,\mathrm{TV}\), and \(R_0 = 90\,\mathrm{TV}\), respectively. The solid lines represent the total emission from cocoons plus the DGE under the \textit{base min} set-up~\citep{Luque:2022buq}. \textbf{Right:} Same as the left panel, but for the outer Galaxy. In both panels, Fermi-LAT and LHAASO data \citep{LHAASO:2024lnz} are included for comparison.}
\label{fig:gamma_different_cutoff}
\end{figure*}

In the main text, we adopt \(R_0 = 20\,\mathrm{TV}\) when estimating the contribution of cocoons to the diffuse background. Here, we examine how varying \(R_0\) influences these results. Figure~\ref{fig:gamma_different_cutoff} illustrates the gamma-ray SEDs of the cocoons for several values of \(R_0\) (from \(10\,\mathrm{TV}\) to \(90\,\mathrm{TV}\), shown as dashed lines), together with their sum when added to the DGE under the \emph{base min} assumption. For \(R_0 \lesssim 10\,\mathrm{TV}\), the cocoon contribution remains negligible. On the other hand, if \(R_0\) is significantly larger than \(20\,\mathrm{TV}\), the total diffuse emission overshoots the low-energy data from LHAASO, thus validating our choice of about \(R_0 \simeq 20\,\mathrm{TV}\).

\section{Impact of different energy dependences of the grammage in the cocoons}
\label{sec:energy_dependence_timescale}

\begin{figure*}[h!]
\centering
\includegraphics[width=0.45\linewidth]{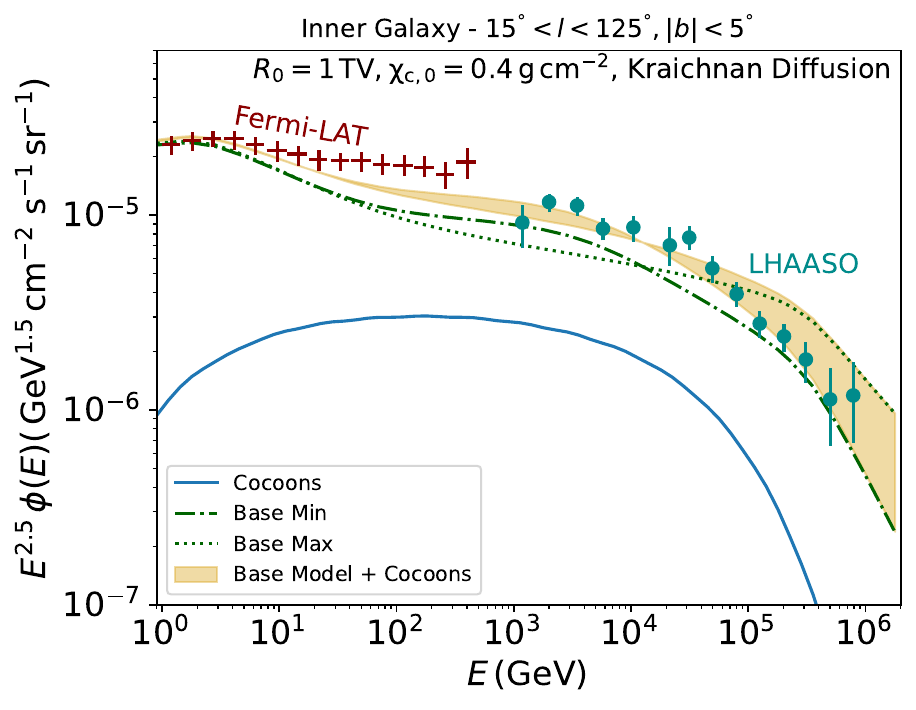}
\includegraphics[width=0.45\linewidth]{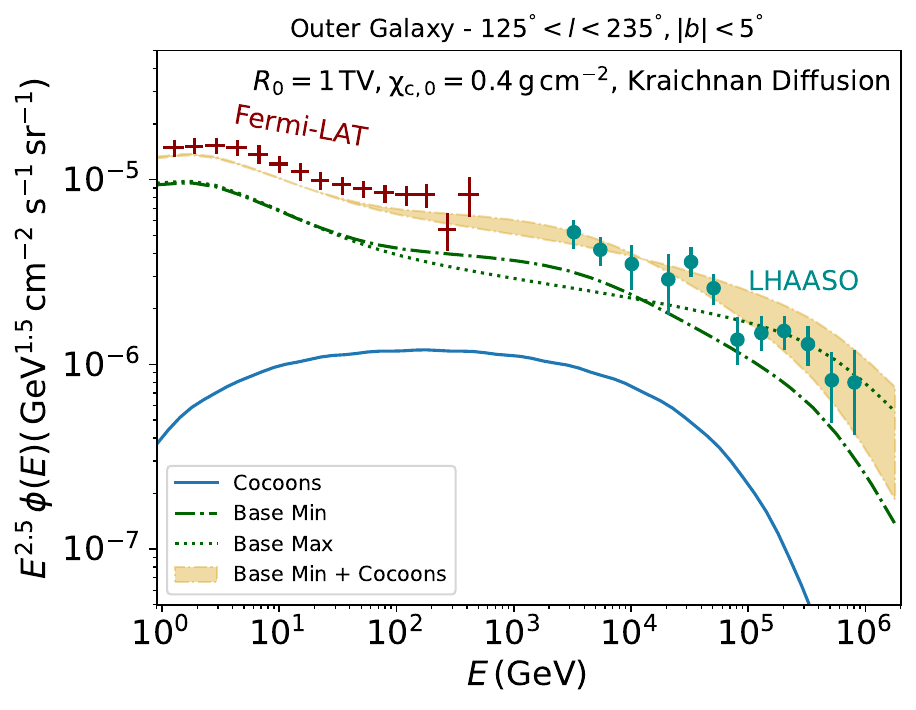}
\includegraphics[width=0.45\linewidth]{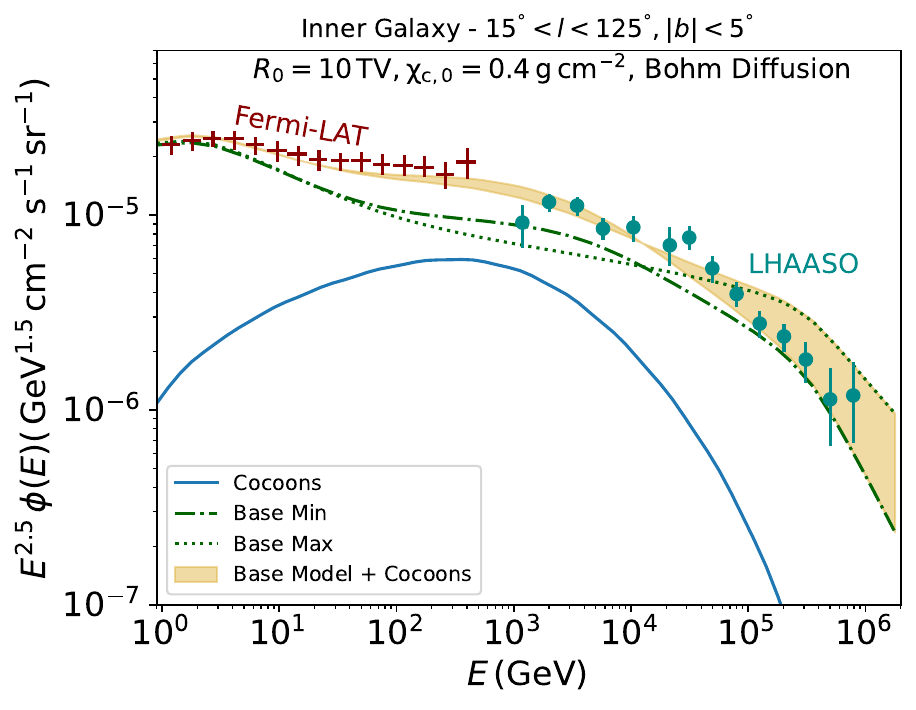}
\includegraphics[width=0.45\linewidth]{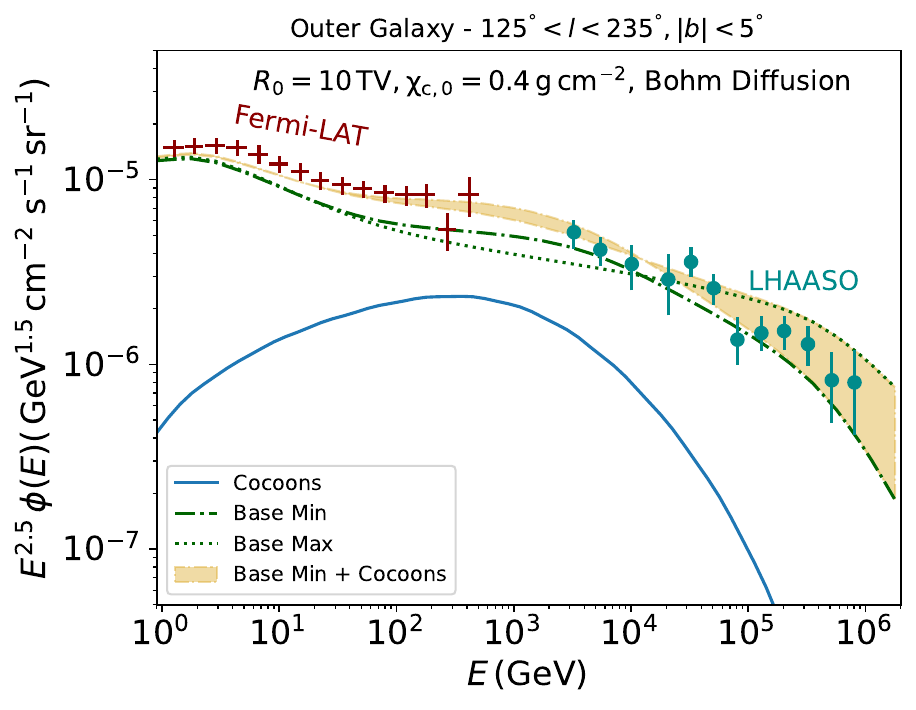}
\caption{\textbf{Left panels}: Gamma-ray SEDs for the inner Galaxy, showing the total emission (DGE + cocoons, golden band) and the cocoon-only contribution (blue line). The predicted SEDs are rescaled to account for the LHAASO mask in the Galactic disc. \textbf{Right panels}: Same as the left panels, but for the outer Galaxy. In the upper panels, we assume Kraichnan diffusion for the confinement time and \(R_0 = 1\,\mathrm{TV}\)), while the lower panels use Bohm diffusion with \(R_0 = 10\,\mathrm{TV}\). The resulting SEDs are compared to Fermi-LAT and LHAASO data from~\cite{LHAASO:2024lnz}.}
\label{fig:gamma_neutrinos_kra}
\end{figure*}

To capture the transition from advection-dominated confinement at low energies to diffusion-dominated confinement at higher energies,  an approach inspired by star clusters as potential hosts of the cocoon scenario,  we adopt the following parametrization:
\begin{equation}\label{eq:timescale_diffusion}
  \chi_c(R) \;=\; \chi_{c,0} \,\biggl[ 1 + \Bigl(\tfrac{R}{R_0}\Bigr)^{\delta} \biggr]^{-1}.
\end{equation}
Here \(\delta\) sets the energy dependence of the diffusion coefficient within the cocoon. As shown in Fig.~\ref{fig:gamma_neutrinos_kra}, the cocoon's secondary emission inherits this energy dependence, most prominently reflected at high energies, where the choice of \(\delta\) drives the spectral shape. In the Kraichnan diffusion case for instance, the cocoon flux can exceed LHAASO’s diffuse measurements, which forces \(R_0 \lesssim 10\,\mathrm{TV}\) to avoid saturating the data. However, such a low \(R_0\) makes it harder to reconcile the cocoon emission with lower-energy Fermi-LAT measurements, as seen in the upper panels of Fig.~\ref{fig:gamma_neutrinos_kra} (where \(R_0 = 1\,\mathrm{TV}\) was assumed). By contrast, Bohm diffusion yields a flux similar to the exponential grammage model up to \(\sim 1\,\mathrm{TeV}\), but produces a comparatively larger emission at higher energies. This can potentially reconcile both Fermi-LAT and LHAASO observations, including the \(\sim 30\text{--}40\,\mathrm{TeV}\) hardening seen in the LHAASO data (see lower panels of Fig.~\ref{fig:gamma_neutrinos_kra}). 

\begin{figure}[t!]
\centering
\includegraphics[width=\linewidth]{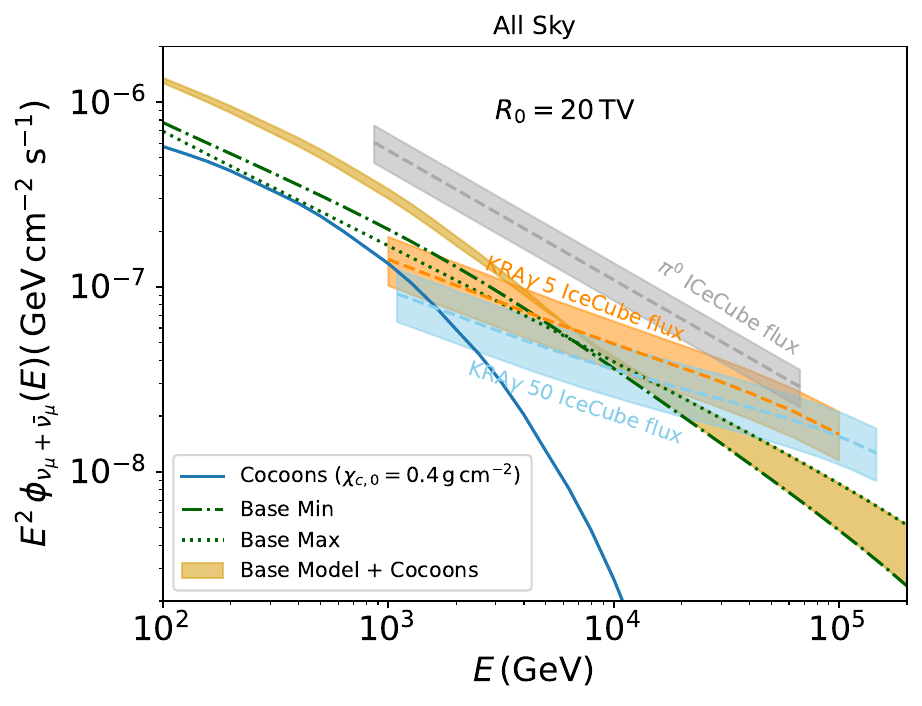}
\caption{All-sky neutrino SEDs as a function of energy, comparing the total emission (DGE + cocoons, shown by the golden band) and the cocoon-only contribution (blue line). These predictions are evaluated against the IceCube flux measurements~\citep{IceCube:2023ame}, derived using the \(\pi^0\) template from~\cite{Fermi-LAT:2016zaq} and the KRA\(\gamma\) 5 and 50 templates described in~\cite{Luque:2022buq}.}\label{fig:all_sky_neutrinos}
\end{figure}

\section{The neutrino contribution of the cocoons}

Cocoons might also contribute to the galactic diffuse neutrino background. Therefore, in Fig.~\ref{fig:all_sky_neutrinos}, we compare our predicted all-sky neutrino flux with the galactic flux measured by the IceCube collaboration using three different spatial templates~\citep{IceCube:2023ame}. Since the DGE + cocoons model provides a different spatial distribution of the signal compared with the models tested by IceCube, in principle the comparison with the IceCube results is not completely self-consistent. On the other hand, the cascade sample used by the collaboration makes the disentanglement between templates rather challenging because of intrinsic angular uncertainty of $\sim 7^{\circ}$ for each event~\citep{Ambrosone:2023hsz}. We note that the cocoons contribute only at $\sim 1-5\, \rm TeV$, making the overall spectrum more complex than the power law used by IceCube, but close to the KRA$\gamma$ results. However, our model does not contribute significantly to the neutrino flux at $100\, \rm TeV$, where all templates appear to converge at approximately $2 \times 10^{-8}\, \rm GeV\, \rm cm^{-2}\, \rm s^{-1}$. Nevertheless, we note that the \textit{base} model is mildly compatible with the measured flux, and we anticipate that future analyses may determine whether this spectral feature truly represents the DGE.

\end{appendix}

\end{document}